\documentclass{article}
\usepackage{spconf,amsmath,epsfig}
\usepackage[utf8]{inputenc}

\title{A review of deep-learning techniques for SAR image restoration}
\name{Loïc Denis$^\dagger$, Emanuele Dalsasso$^\ddagger$ and Florence
Tupin$^\ddagger$\thanks{The
authors would like to thank the French
space agency CNES for funding this work.}}
\address{$^\dagger$Univ Lyon, UJM-Saint-Etienne, CNRS,
Institut d Optique Graduate School,\\ Laboratoire Hubert Curien UMR
5516,
F-42023, SAINT-ETIENNE, France\\[1ex]
 $^\ddagger$LTCI, Telecom Paris, Institut Polytechnique de Paris, 91120
 Palaiseau, France}

\begin{document}
%
\maketitle
\begin{abstract}
The speckle phenomenon remains a major hurdle for the analysis of SAR images. The development of speckle reduction methods closely follows methodological
progress in the field of image restoration.
The advent of deep neural networks has offered new ways to tackle this
longstanding problem. Deep learning for speckle reduction is a very active
research topic and already shows restoration performances that exceed that of
the previous generations of methods based on the concepts of patches,
sparsity, wavelet transform or total variation minimization.

The objective of this paper is to give an overview of the most recent works
and point the main research directions and current challenges of deep learning
for SAR image restoration.
\end{abstract}
\section{Introduction}
\label{sec:intro}

Speckle phenomenon arises due to the coherent summation of echoes produced by
elementary scatterers that project into the same SAR pixel. Mitigating the
strong fluctuations of speckle has been a major issue since the beginnings of
SAR imaging.

Multilooking, i.e., averaging SAR intensities in a spatial window around the
pixel of interest, reduces speckle fluctuations at the cost of a dramatic
resolution loss. More subtle approaches have thus been proposed to prevent
from blurring structures with very different reflectivities:
\emph{pixel-selection} methods restrict the average to intensities close to
that of the current pixel, \emph{window-based} methods adapt the shape of the
window (by locally selecting a window among a set of oriented windows, or by
region growing), \emph{patch-based} methods compare patches to identify
(possibly disconnected) pixels with similar neighborhoods,
\emph{transform-based} techniques apply a transform (such the wavelet
transform) to separate noise from the useful signal, \emph{regularization} or
\emph{variational} methods minimize a cost function that expresses a tradeoff
between the proximity to the speckled observation and spatial smoothness
properties. Deep learning is a much more recent approach to speckle reduction.
The data-driven nature of this approach offers an improved flexibility and the
ability to capture a wide variety of features observed in SAR images
(point-like scatterers, lines, curves, textures). In the following we describe
how deep learning methods are designed and describe the main challenges of
this
quickly evolving research topic.

\section{Key ingredients of a deep learning approach for SAR despeckling}

\newcommand{\One}{\protect\raisebox{-.2ex}{\protect\includegraphics[width=2ex]{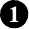}}}
\newcommand{\Two}{\protect\raisebox{-.2ex}{\protect\includegraphics[width=2ex]{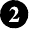}}}
\newcommand{\Three}{\protect\raisebox{-.2ex}{\protect\includegraphics[width=2ex]{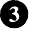}}}
\begin{figure*}[t]
    \centerline{\includegraphics[width=\textwidth]{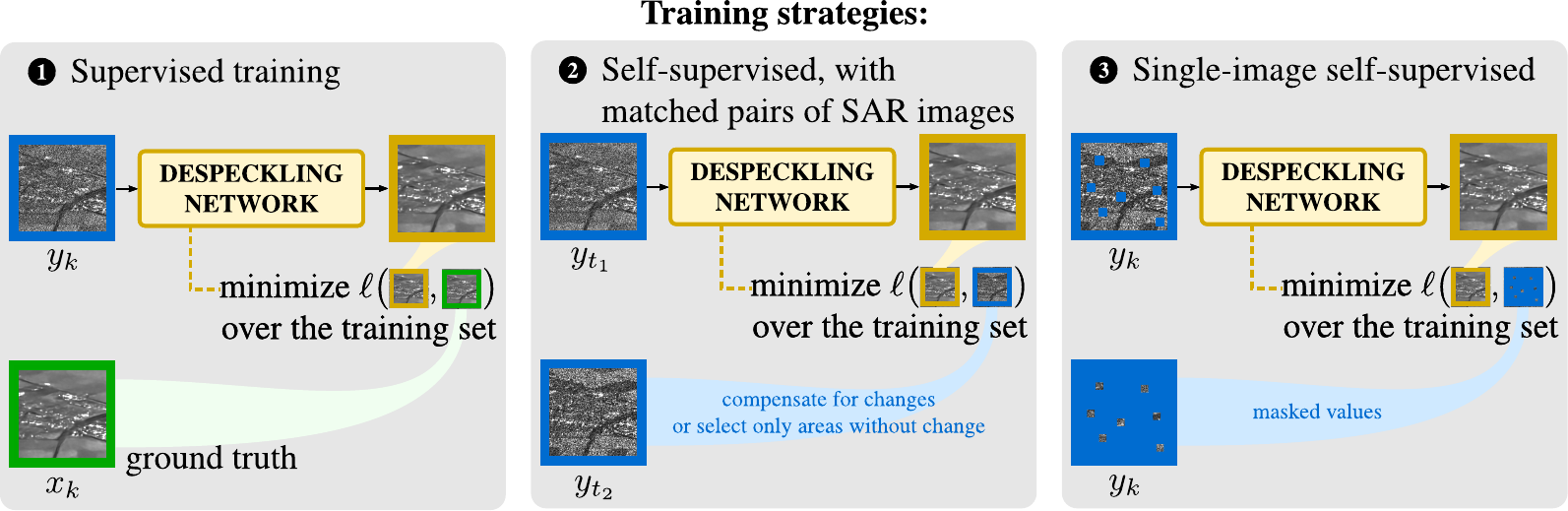}}\vspace*{-1ex}
    \caption{Three training strategies have been explored in the literature:
        \One{} supervised training, using ground-truth images that match the
        speckled images provided as input to the network; \Two{}
        self-supervised
        training, using co-registered pairs of SAR images captured at
        different
        dates; \Three{} self-supervised training, using single images and a
        masking strategy: the network is trained to correctly infer the masked
        pixels of the input image.}
    \label{fig:supervision}
\end{figure*}

\begin{figure*}[t]
    \centering\begin{tabular}{ccc}
        &\includegraphics[width=.2\textwidth]{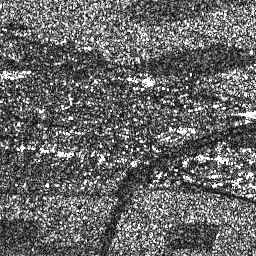}\\
        &(a) Sentinel-1 (SLC)\\[1ex]
        \includegraphics[width=.2\textwidth]{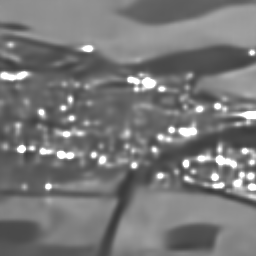}&
        \includegraphics[width=.2\textwidth]{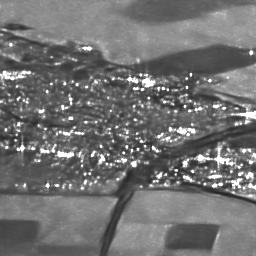}&
        \includegraphics[width=.2\textwidth]{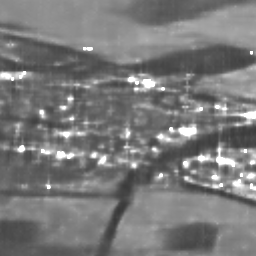}\\
        (b) restoration with SARCNN &(c) restoration with SAR2SAR & (d)
        restoration with speckle2void.
    \end{tabular}     \vspace*{-2ex}
    %
    \caption{Restoration of the single-look Sentinel-1 image shown in (a) with
        deep-learning methods illustrative of the 3 training strategies shown
        in Fig.\ref{fig:supervision}: (b) SAR2SAR \cite{chierchia2017sar} uses
        a
        {\bf supervised training}
        strategy
        (here, the training is performed on synthetic speckle and the
        Sentinel-1
        image is downsampled by a factor 2 to limit speckle correlation, see
        \cite{dalsasso2020sar}); (c)
        restoration with SAR2SAR \cite{dalsasso2020sar2sar}, a network trained
        with a {\bf self-supervised
            approach with pairs} of Sentinel-1 images of the same area
            captured at
        different dates; (d) restoration with the
        {\bf single-image self-supervised} method speckle2void
        \cite{molini2020speckle2void}.}
    \label{fig:illus_results}
\end{figure*}

\subsection{Building a training set}
A first but crucial step to design a deep learning method for speckle
reduction is the choice of a training strategy. The most conventional approach
to train a network is \emph{supervised training}
(Fig.\ref{fig:supervision}, block \One). This strategy requires the building
of a training set with pairs of speckled~/ speckle-free images. Such pairs can
be obtained by generating simulated speckle from a ground-truth image. It is
however difficult to obtain such speckle-free images. The main approach
consists in reducing speckle fluctuations by temporally averaging images from
a long time series. A
major limitation of numerically generated speckle, though, is that it generally
neglects speckle correlations. The shift between the speckled images used
during training and the real images used at test time produces strong
artifacts unless adaptations are done, such as image downsampling
\cite{dalsasso2020handle}, or training on regions of real images carefully
selected to reject any area that changed during the time series
\cite{chierchia2017sar}. To prevent these limitations, \emph{self-supervised
strategies} use only speckle-corrupted images in the training phase. Pairs of
co-registered SAR images obtained at two different dates (chosen so that
speckle is
temporally decorrelated between the images) can be used to drive the network
to predict an estimate from the first image that is as close as possible to
the second image (Fig.\ref{fig:supervision}, block \Two). Single-image
self-supervision introduces a form of masking: the network accesses only
unmasked values and is asked to guess the masked values
(Fig.\ref{fig:supervision}, block \Three). Given the random nature of speckle
phenomenon, the best guess for the network is the underlying reflectivity
(i.e., the noiseless value at the masked pixel).

\subsection{Choosing a network architecture}
There is a wide variety of network architectures available for image
denoising. Two kinds of
networks are generally used for SAR despeckling: (i) the convolutional
structure of DnCNN \cite{zhang2017beyond} (obtained by stacking 15 to 20
layers formed by convolutions, possibly with dilation \cite{zhang2018learning}, batch normalization and a ReLU activation
function), trained in a residual fashion, and (ii) the U-Net
\cite{ronneberger2015u} (originally used for image segmentation, that takes
the form of a particular auto-encoder with skip-connections).

\subsection{Handling the high dynamic range of SAR images}

Due to the physics of SAR imaging, the dynamic range between echoes produced
by weakly scattering surfaces and the very strong returns generated by
trihedral structures typically spans several orders of magnitude.
Normalization and compression of the range of SAR image intensities is a
crucial step: it strongly reduces the risk of falling outside the domain
covered during the training phase of the network. Many works apply a logarithm
transform to the SAR intensities before the deep neural network. This has two
beneficial effects: it compresses the range of input values (so that it is
much less likely to find strongly out-of-range values at test time) and it
stabilizes
the variance of speckle fluctuations (which may simplify despeckling). When
SAR images are processed by the network in the original domain (i.e., without
log-transform), the largest values are typically clipped to reduce the dynamic
range, see for example \cite{molini2020speckle2void}.

\subsection{Selecting a loss function}

The most widely used loss function for regression is the squared $\ell_2$
norm. To reduce the impact of the training samples that are poorly modeled, an
$\ell_1$ norm can be preferred. Total variation is sometimes considered as an
additional term
to penalize oscillations and thus limit the apparition of artifacts when
applied to images that differ from the distribution of images considered
during training (e.g., when speckle is spatially correlated at test time)
\cite{dalsasso2020handle}. Loss terms that enforce a good fit with the
theoretical distribution of speckle have also been recently considered
\cite{vitale2020multi}.
Perceptual losses can be used in supervised
training strategies to give more weight to artifacts that may be interpreted
as visual clues of meaningful content in the image. Generative Adversarial
Networks (GANs) can be used to train a discriminator whose aim is to recognize
restored images based on some artifacts of the restoration technique. Training
the restoration network to fool the discriminator is then a way to obtain more
plausible restoration results, at the cost of increasing the risk of also
fooling the human by adding fake content that looks realistic \cite{wang2017generative}.

Self-supervised training strategies require adapted loss functions. In the
case of self-supervision with matched pairs of SAR images, it is important to
compensate for changes that occurred between the two dates
\cite{dalsasso2020sar2sar}. Single-image
self-supervision requires to limit the computation of the loss to the masked
pixels, or the use of a specific network architecture that prevents the
receptive field to contain the central pixel \cite{molini2020speckle2void}.

\section{Current challenges and trends}

\subsection{Self-supervision}

In remote sensing, huge amounts of images are available but ground truths are
scarce and costly to produce. Numerical simulations only imperfectly reproduce
the complexity of actual systems. The development of learning strategies that
rely solely on actual observations is thus very appealing. Specific challenges
face these strategies, however, such as the compensation of temporal changes
(when co-registered image pairs acquired at different dates are considered) or
the correlation of speckle (in particular for masking approaches).

\subsection{Extensions to polarimetric and/or interferometric SAR}
Most deep learning approaches for speckle reduction focused on the case of
intensity images. Multi-channel complex-valued SAR images, as in SAR
polarimetry or in SAR interferometry, raise other challenges. Polarimetric and
interferometric information are encoded in complex-valued covariance matrices.
Restricting the estimated matrices to the cone of positive definite covariance
matrices requires an adequate design of the learning strategy and/or of the
network. Due to the increase of the dimensionality of the data and of the
unknowns, the learning task becomes more complex and it is expected that many
more training samples are required to capture all spatial and
polarimetric/interferometric configurations during the learning phase.

\begin{figure}[t]
\centerline{\includegraphics[width=\columnwidth]{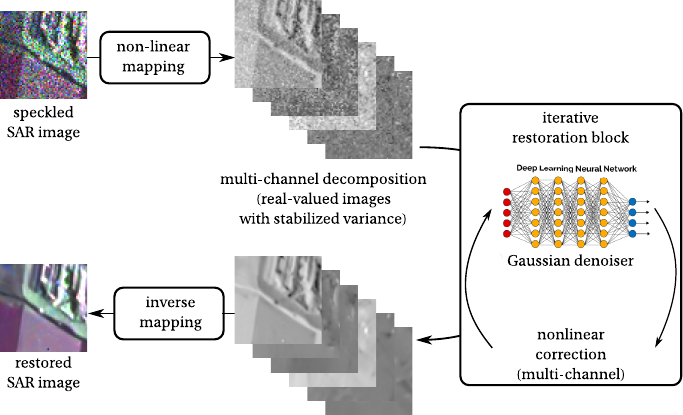}}
\caption{MuLog \cite{deledalle2017mulog} is one of the first approaches to
apply
deep neural networks to speckle reduction in polarimetric and interferometric
SAR restoration. It works in a transformed domain in which complex-valued
polarimetric and/or interferometric matrices are decomposed into real-valued
channels with an approximately stabilized variance. In this domain, a deep
neural network is applied iteratively until the channels are restored.
Extending deep learning methods to polarimetric and/or interferometric SAR
data is
a hot topic.}
\label{fig:mulog}
\end{figure}

A notable approach to address these issues consists in applying a plug-in ADMM
strategy to account for the statistics of speckle in polarimetric and
interferometric SAR imaging \cite{deledalle2017mulog}. By decomposing the SAR
images into almost
independent channels, deep neural networks can be readily applied, see
Fig.\ref{fig:mulog} and
\cite{deledalle2018mulog}.

\subsection{Extension to time series}
Satellite constellations such as ESA's Sentinel-1 provide very long time
series. The frequent revisit time and the temporal decorrelation of speckle
offer the potential of very effective speckle suppression by (spatio)-temporal
filtering. Versatile networks able to process temporal stacks of various size
would be of great value to analyze these images.

\subsection{Understanding and characterizing the restoration results}
A limitation of deep learning methods is their lack of explainablity: due to
the highly non-linear nature of the networks and their numerous parameters, it
is very hard to grasp how a network produced a given result and to
characterize the different artifacts that may be produced at test time. An
approach to improve the explainability of deep learning methods is to combine
them with more traditionnal processing techniques such as patch-based methods
\cite{denis2019patches,cozzolino2020nonlocal}.


\section{Conclusion}

SAR image restoration with deep neural networks is an extremely active
research area, with very convincing results and several open research
directions. The limited space of this paper was unsufficient to adequately
cite the quickly growing literature on the subject. We focused on providing a
broad view on the key elements of deep learning techniques for speckle
reduction and invite the interested reader to refer to much more extensive
reviews such as \cite{zhu2020deep} and \cite{fracastoro2020deep}.


\bibliographystyle{IEEEbib}
\bibliography{refs}

\end{document}